\documentclass{appolb}
\usepackage{epsfig} 
\usepackage[english]{babel} 
\usepackage{amsmath}
\usepackage{amsfonts}

\def\ds{\displaystyle} 
 
\title{Atomic Nuclei with Tetrahedral and Octahedral Symmetries\thanks
      {Presented at 
       XXXVII Zakopane School of Physics, Zakopane, 3 - 10 Sept. 2002 .}
}
 
\author{J.~Dudek$^1$,  A.~G\'o\'zd\'z$^{\,1,2}$ and N. Schunck$^1$ 
\address{{\it $^1$Institut de Recherches Subatomiques,
              IN$_2$P$_3$-CNRS/Universit\'e Louis Pasteur \\
              F-67037 Strasbourg Cedex 2, France}}
\address{{\it $^2$Katedra Fizyki Teoretycznej, 
              Uniwersytet Marii Curie-Sk\l{}odowskiej,}\\
              {\it PL-20031 Lublin, Poland}}
}
\begin{document}

%
%
\maketitle

\begin{abstract} 
   
We present possible manifestations of octahedral and tetrahedral symmetries in
nuclei. These symmetries are associated with the $O_h^D$ and $T_d^D$ double
point groups. Both of them have very characteristic finger-prints in terms of
the nucleonic level properties - unique in the Fermionic universe. The
tetrahedral symmetry leads to the four-fold degeneracies in the nucleonic
spectra; it {\em does not preserve} the parity. The octahedral symmetry leads
to the four-fold degeneracies in the nucleonic spectra as well but it {\em does
preserve} the parity.

Microscopic predictions have been obtained using  mean-field theory based on
the relativistic equations and confirmed by using 'traditional' Schr\"odinger
equation formalism. Calculations are performed in multidimensional 
deformation spaces using newly designed algorithms.  

We discuss some experimental fingerprints of the hypothetical new symmetries
and possibilities of their verification through experiments.

\end{abstract} 


 
\section{Introduction}
\label{Sect:01}
 

The phenomenon of the shape coexistence in nuclei is related to one of those
'intuitive' mechanisms that can be relatively easily imagined in terms of
classical physics and geometry. This is perhaps one of the reasons why a
conceptual progress in this important sub-field of nuclear structure physics
has been relatively slow - although important successes such as finding an
evidence for 'prolate-spherical-oblate' shape coexistence or for a coexistence
between the super-deformed and normal-deformed nuclear configurations - have
been achieved. The examples of the shape coexistence just mentioned are related
directly to fundamental symmetries of the nuclear mean-field: (pseudo) SU$_3$
and the so-called  pseudo-spin symmetries (cf. e.g. Ref.~\cite{JDu87},
\cite{JDu02}). Studying the symmetries that will be discussed in this paper may
shed some light on yet another microscopic mechanism possibly present in
nuclei: the spontaneous symmetry breaking leading to {\em unusually high
single-nucleon degeneracies} that may appear in {\em deformed} nuclei. 

The most common, one may think, triaxial-ellipsoid nuclear shapes have for a
long time {\em not} received as thorough an attention from the experimental
point of view as they should have - even though in many models, such as
cranking model or mean-field theory based models, the tri-axiality and the
so-called $\gamma$-deformation play an important role. Only recently an attempt
to observe the quantum wobbling mechanism in nuclei has forced considering a
simultaneous combination of several experimental manifestations  of the
non-axial shapes in nuclei. To our knowledge, there has been so far no
experimental effort undertaken, in terms of searches for an evidence of
non-axially symmetric nuclear shapes, with the exception of the ellipsoidal
ones. 

In terms of the conceptual progress, the so-called $\mathcal{C}_4$-symmetry 
hypothesis is particularly worthwhile mentioning, Ref.~\cite{IHB96}. Although
no consistent evidence of its presence in experimental results on
super-deformed nuclei exists so far (and rather numerous arguments against), no
systematic search, neither theoretical nor experimental in terms of {\em
normally deformed} nuclei has ever been undertaken, and it remains to be seen
which nuclei can possibly built-up  $\mathcal{C}_4$-symmetric configurations
with the elongation that are not very different from their ground-state
elongation.

Recently, an idea originally proposed in Ref.~\cite{LiD96} has been re-analyzed
in terms of the possible presence of the pyramid-like (tetrahedral) shapes in
nuclei, Ref.~\cite{JDA02}, with a conclusion that extremely strong nuclear
shell effects leading to a tetrahedral symmetry may exist in nature on a
sub-atomic level. In this presentation we would like to address a slightly more
general problem of possible existence in nuclei of both {\em octahedral} and
{\em tetrahedral} symmetries; these symmetries are mathematically related
but cause very different physical implications.

Octahedral and tetrahedral symmetries are characterized by a relatively large
number of symmetry elements. Compared to classical $D_2$-symmetry group that is
composed of 4 elements (three rotations through an angle of $\pi$ about three
mutually perpendicular axes plus the identity transformation) and characterizes
a family of tri-axially deformed nuclei, a group of symmetry of a classical
tetrahedron, $T_d$, contains 24 symmetry elements and that of a classical
octahedron, $O_h$, 48 symmetry elements\footnote{On the level of symmetries of
the Schr\"odinger equation for the nucleons (fermions) the classical symmetry
groups need to be replaced by the so-called double (or spinor) groups that
contain a double set of symmetry elements. In the mentioned groups this means
8, 48 and 96 symmetry elements for the double $D^D_2$, $T^D_d$ and $O^D_h$
groups, respectively.}. As a result of such a high degree of symmetry, the
$T^D_d$ or $O^D_h$ invariance implies an unusually high degeneracy of the
single particle states - eigen-solutions to the Schr\"odinger equation. More
precisely, the double tetrahedral, $T^D_d$-symmetry group generates two
two-dimensional and one four-dimensional irreducible representations. This fact
manifests itself through double and quadruple degeneracies of the
single-nucleonic levels - an unusual situation given the fact that so far, for
the deformed nuclei, only the double (i.e. Kramers) degeneracies of the
single-nucleonic levels have been considered.

The octahedral double point group, $O^D_h$, contains an inversion among its
symmetry elements with the consequence that the parity of single-nucleonic
levels is preserved by the solutions to the octahedrally-symmetric
hamiltonians. In this case we find six irreducible representations, three of
them characterized by the positive parity of the underlying single-particle
states and three other by the negative parity. Within each of the two parities
we find two two-dimensional and one four-dimensional irreducible
representations and it follows that the corresponding levels can be occupied by
up to two- and up to four nucleons, respectively.

%
\section{Symmetries of the Nuclear Mean-Field}
\label{Sect:02}
%

In this Section we are going to summarize the mathematical concepts underlying
the present study. This summary will be followed by a few illustrations of the
discussed principles in the case of the nuclear $T^D_d$ and $O^D_h$ symmetries.

%
\subsection{General Aspects of Discrete Symmetries in Multi-Fermion Systems}
\label{Sect:02.1}
%
 
We consider a deformed mean-field nuclear hamiltonian; the corresponding 
operator can always be written down in the form
\begin{equation}
      \hat{\mathcal{H}} 
       = 
      \hat{\mathcal{H}}(\vec{r}, \vec{p}, \vec{s};\hat{\alpha})
                                                                  \label{eqn01}
\end{equation}
where $\vec{r}$, $\vec{p}$ and $\vec{s}$ are the position, linear momentum and
spin operators, respectively, and where $\hat{\alpha} \leftrightarrow \{
\alpha_{\lambda,\mu} \}$ represents an ensemble of all parameters that define
nuclear shapes; here we are using the multipole deformation parameters that are
particularly well suited for analyzing the point-group symmetry properties.

Consider a group $\mathcal{G}$ with the symmetry operators (group elements)
\begin{equation}
      \{ {\hat{\mathcal{O}}}_1, 
         {\hat{\mathcal{O}}}_2, \; \ldots 
         {\hat{\mathcal{O}}}_f \}
      \Leftrightarrow
      \mathcal{G}. 
                                                                  \label{eqn02}
\end{equation}
Assuming that $\mathcal{G}$ is the group of symmetry of hamiltonian 
$\hat{\mathcal{H}}$ implies that all elements of the group commute with
$\hat{\mathcal{H}}$:
\begin{equation}
      [ \hat{\mathcal{H}}, {\hat{\mathcal{O}}}_k ]
      =0
      \quad {\rm with} \quad
      k=1,2,\; \ldots f \; .
                                                                  \label{eqn03}
\end{equation}
Of course operators $\hat{\mathcal{O}}_k$ may, but do not need to commute among 
themselves. Suppose that the group in question has irreducible representations
(irreps)
\begin{equation}
      \{ {\mathcal{R}}_1, {\mathcal{R}}_2, \; \ldots {\mathcal{R}}_r \} \; .
                                                                  \label{eqn04}
\end{equation}
(The reader unfamiliar with the terminology used in group theory does not need
to know at this point more than the fact that the irreps can be characterized
by their dimensions.) Suppose that the irreps in questions have dimensions
\begin{equation}
      \{\, d_1, d_2, \; \ldots \; d_r \, \} \; 
                                                                  \label{eqn05}
\end{equation}
respectively. Then the eigenvalues $\varepsilon_{\nu}$ of the problem
\begin{equation}
      \hat{\mathcal{H}} \,\Psi_{\nu}
      =
      \varepsilon_{\nu} \,\Psi_{\nu}
                                                                  \label{eqn06}
\end{equation}
appear in multiplets: $d_1$-fold degenerate, $d_2$-fold degenerate, $\ldots$ 
etc. 

The point groups of interest for us in nuclear physics applications differ from
those usually discussed in crystallography. Since the eigenstates of the
problem in Eq.~(\ref{eqn06}) are spinors it follows that all $360^o$
space-rotations {\em must not give} the identity, $\mathcal{I}$, but  rather
$-\mathcal{I}$ (change in phase). The classical crystallographic point groups
'adapted' to provide this feature are called double or spinor point groups and
their names are written with the superscript $D$, as seen already above. Among
32 standard point groups usually  considered in quantum mechanics applications
(cf. Ref.~\cite{32grp} for a detailed presentation) there are only two that are
of interest for the present study: the tetrahedral and the octahedral double
point groups, $T^D_d$ and $O^D_h$, respectively. The important physical reason
for that interest is that among the corresponding irreps there are some with
dimensions $d=4$; all other double point-groups of possible interest in
subatomic physics generate exclusively the double degeneracies at most.

%
\subsection{Four-Fold Degeneracies of Nucleonic Levels and Energy Gaps}
\label{Sect:02.2}
%

The degeneracies of single-nucleonic levels are related to the symmetries of
the underlying potential. They may imply a presence of strong gaps in the
single particle spectra and thus play an important role in stabilizing  certain
nuclear configurations. Indeed, if a large energy gap appears at the Fermi
level of a given nucleus, in order to excite (i.e. destabilize) the
corresponding configuration for instance through bombarding with external
particles, there will be a large energy necessary as compared to the situations
where the gaps are small. This mechanism is very well known in spherical nuclei
in which the 'magnetic' [(2j+1)-fold] degeneracy of the nucleonic levels gives
rise to the large ('magic') gaps and implies indeed a strong increase in
stability of the corresponding nuclei, like e.g. in $^{208}$Pb.

What has been until recently unknown is that the single-particle spectra in 
deformed nuclei may generate gaps as large as those in the spherical ones~(!)
and that apparently the $T^D_d$ and/or $O^D_h$ symmetries play an important
role there. The quantitative predictions related to the strong shell-gaps
presented below, can be qualitatively understood as follows.  

The property of saturation of the nuclear forces leads, among others, to a
relatively weak dependence on the proton and neutron numbers of the depth of
the mean nuclear potential. In fact in several nuclear physics considerations
this weak dependence has been neglected altogether assuming that the potential
depth remains constant (this will of course {\em not} be the case in the
present study when performing the realistic calculations and the argument is
brought here for the sake of a qualitative consideration only). At a constant
depth of the potential, an enhanced appearance of single-particle gaps in the
spectra is more likely if the increased degeneracy of levels is allowed. The
above statement is based on 'empirical' knowledge and has no rigorous
mathematical foundation\footnote{In fact the single-particle spectra that
differ in terms of the level degeneracies, if obtained with the mean-field
potentials of (nearly) constant depths will differ in terms of the average
level spacing: the higher the degeneracies - the larger the average level
spacing. For the realistic nuclear hamiltonians there is no general theorem
that allows to predict the presence (or absence), of the large gaps in the
single particle spectra although some aspects like e.g. the influence of the
spin-orbit interaction potential on such gaps are, to some extent,
predictable.}.   

%
\section{Octahedral and Tetrahedral Symmetries: Shapes}
\label{Sect:03}
%

One of the important mathematical aspects of working with the octahedral and
tetrahedral symmetries is related to modeling of these symmetries with the
help of spherical harmonics when parameterizing the nuclear surface~$\Sigma$:
\begin{equation}
      \Sigma: \quad \mathcal{R}(\vartheta , \varphi; \hat{\alpha})
      =
      R_0 \; c(\hat{\alpha}) \;
      [1 
      + 
      \ds\sum_{\lambda = 2}^{\lambda_{max}} \sum_{\mu = -\lambda}^{\lambda}
      \alpha_{\lambda , \mu} \; Y_{\lambda , \mu}(\vartheta, \varphi) ] \; .
                                                                  \label{eqn07}
\end{equation}
Above, $\hat{\alpha} \equiv \{ \alpha_{\lambda,\mu}; \; \lambda = 2, 3, \;
\ldots \lambda_{max}; \; \mu= -\lambda, \; -\lambda+1,\dots \; +\lambda\}$,
$R_0$ is the nuclear radius parameter,, and $c(\hat{\alpha})$, a function whose
role is to insure that the nuclear volume remains constant, independent of the
deformation.

%
\subsection{Octahedral Deformations}
\label{Sect:03.01}
%

We can demonstrate that there exist special combinations of spherical harmonics
that can be used as a basis for surfaces with octahedral symmetry. The lowest
order of the octahedral deformation, {\em called by convention the first}, is
characterized by the fourth rang spherical harmonics. By introducing a single
parameter $o_1$ we must have in this case
\begin{equation}
      \alpha_{40} \equiv + o_1;
      \quad
      \alpha_{4,\pm 4}\equiv + \sqrt{\frac{5}{14}} \cdot o_1 \; ,
                                                                  \label{eqn08}
\end{equation}
i.e. three hexadecapole deformation parameters must contribute simultaneously
and with proportions $\sqrt{5/14}$ fixed by the octahedral symmetry 
\vspace{-0.5truecm}
\begin{figure}[h] 
\begin{center} 
\includegraphics[scale=0.30]{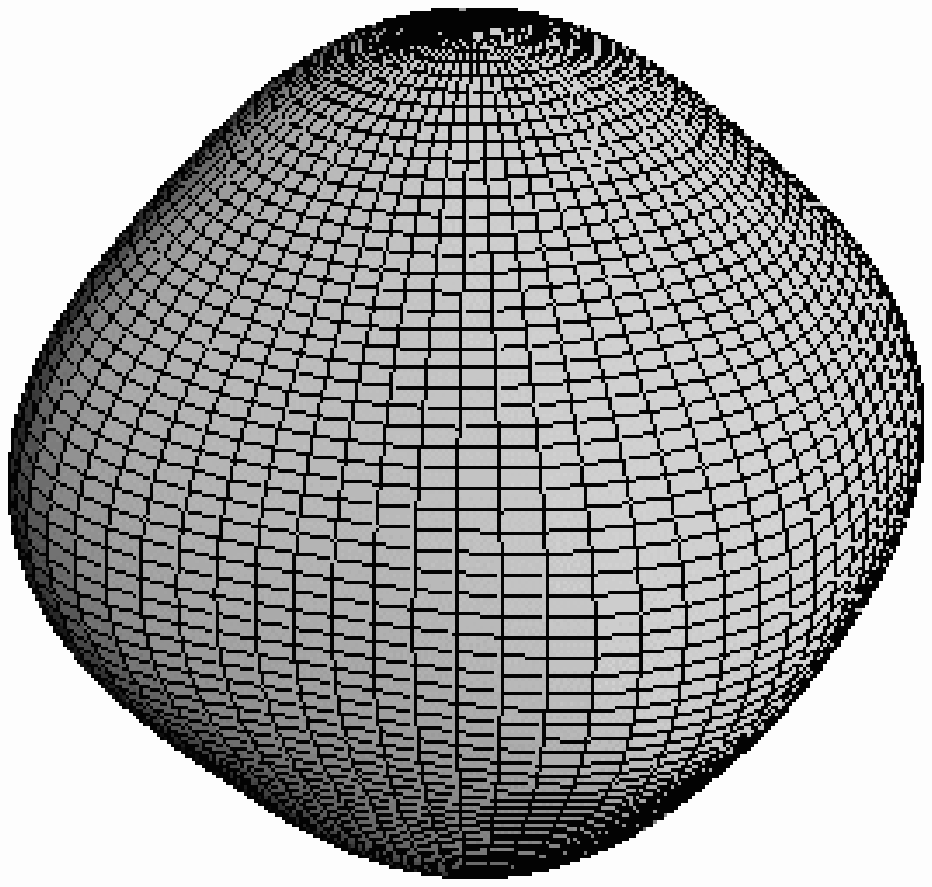}
\includegraphics[scale=0.25]{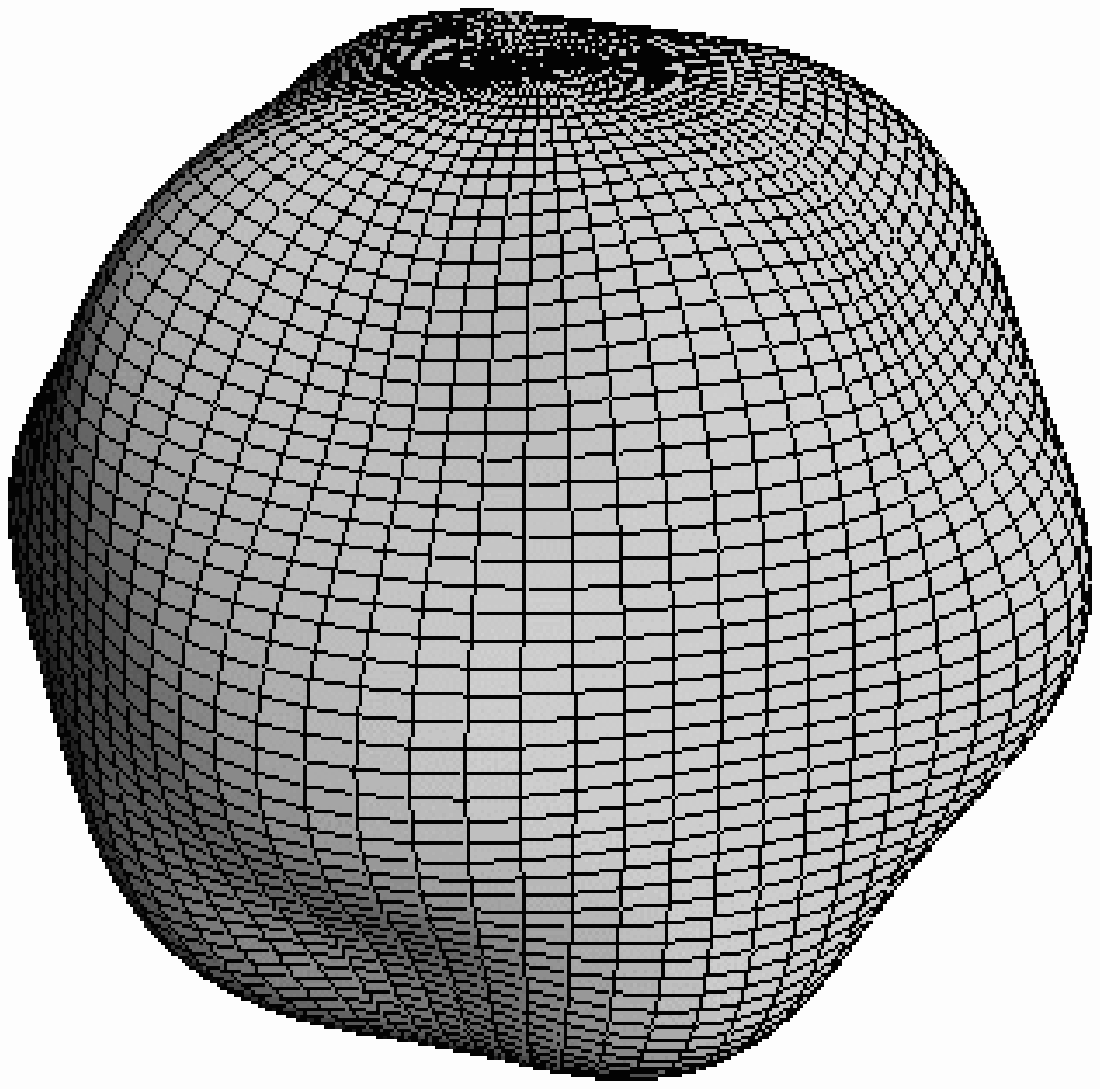}
\vspace{-0.5truecm}
\caption{\small 
         Comparison of two octahedrally deformed nuclei. 
         Left: octahedral deformation of the first order, o$_1$=0.10; 
         right: octahedral deformation of the second order, o$_2$=0.04 .
}                                  
\label{fig01}                 
\end{center} 
\end{figure}
requirement. No deformation with $\lambda = 5$ is allowed and the next possible
are deformations with spherical harmonics of $\lambda=6$. Similarly, we
introduce one single parameter, $o_2$, with the help of which the next allowed
octahedral deformation, called {\em of the second order}, and depending on the
6$^{th}$ rang spherical harmonics can be defined. We must have:
\begin{equation}
      \alpha_{60} \equiv + o_2;
      \quad
      \alpha_{6,\pm 4}\equiv - \sqrt{\frac{7}{2}} \cdot o_2 \; .
                                                                  \label{eqn09}
\end{equation}
The {\em third order} octahedral deformation is characterized by the 8$^{th}$
rang spherical harmonics and we can demonstrate that it can be defined with the
help of a single parameter, $o_3$, where:
\begin{equation}
      \alpha_{80} \equiv + o_3;
      \quad
      \alpha_{8,\pm 4}
      \equiv  
      \sqrt{\frac{28}{198}} \cdot o_3;
      \quad
      \alpha_{8,\pm 8}
      \equiv 
      \sqrt{\frac{65}{198}} \cdot o_3 \; .
                                                                  \label{eqn10}
\end{equation}

Of course the basis of the octahedrally deformed surfaces is infinite, but the
increasing order of the octahedral deformations implies immediately a twice as
fast increase in the rang of the underlying multipole deformations; the
possibility of having such a situation in real nuclei is unlikely and the
expansion series can be cut off quickly. 

%
\subsection{Tetrahedral Deformations}
\label{Sect:03.02}
%

In a similar fashion, the tetrahedral deformation basis can be introduced in
terms of the standard spherical harmonics. The first order tetrahedral
deformation, $t_1$, is characterized by a single octupole deformation with
$\lambda=3$ and $\mu = 2$  and we have
\begin{equation}
      \alpha_{3, \pm 2} \equiv  t_1
                                                                  \label{eqn11}
\end{equation}
\begin{figure}[h] 
\begin{center} 
\vspace{-0.5truecm}
\includegraphics[scale=0.25]{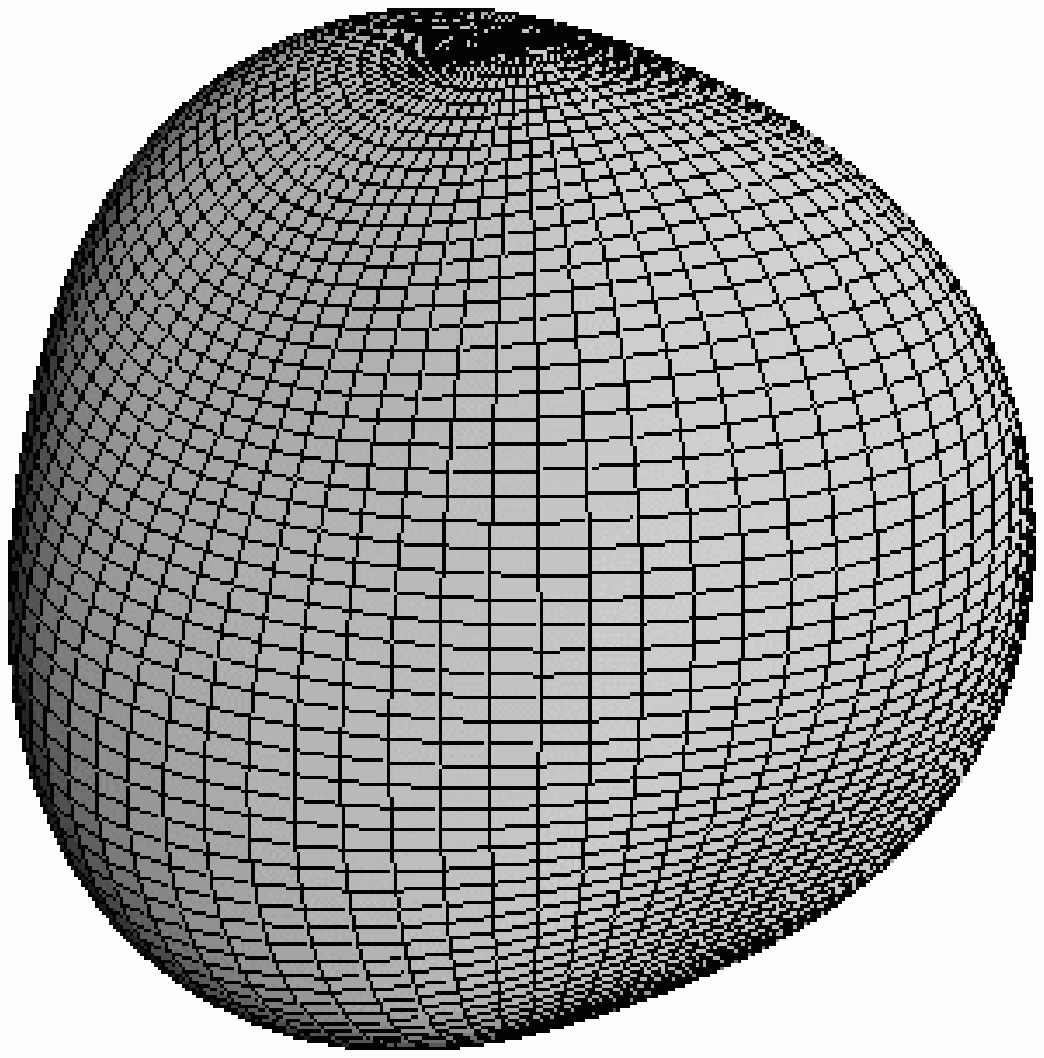}
\includegraphics[scale=0.25]{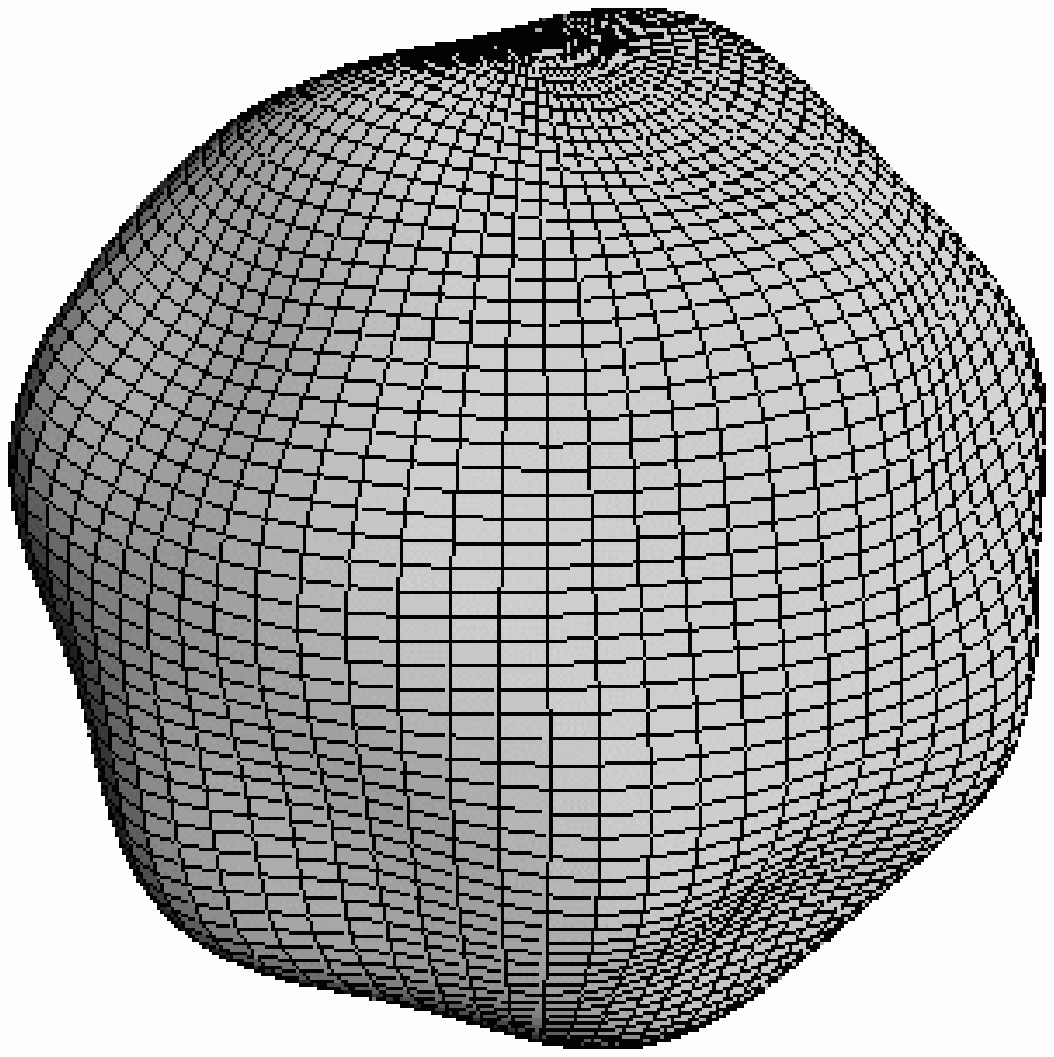}
\vspace{-0.75truecm}
\caption{\small 
         Comparison of two tetrahedrally deformed nuclei. 
         Left: tetrahedral deformation of the first order, t$_1$=0.15; 
         right: tetrahedral deformation of the second order, t$_2$=0.05 .
}                                  
\label{fig02}                 
\end{center} 
\end{figure}
The second order tetrahedral deformation, $t_2$, is characterized by 
multipolarity $\lambda=7$ (observe that the multipoles with $\lambda
=4,5$ and 6 are not allowed at all by the symmetry studied) and we have
\begin{equation}
      \alpha_{7, \pm 2} \equiv t_2 \, ;
      \qquad
      \alpha_{7, \pm 6} \equiv - \sqrt{\frac{11}{13}} \cdot t_2 \; . 
                                                                  \label{eqn12}
\end{equation}
The third order tetrahedral deformation, $t_3$, is characterized by $\lambda=9$ 
and by definition we must have:
\begin{equation}
      \alpha_{9, \pm 2} \equiv t_3;
      \quad
      \alpha_{9, \pm 6} \equiv + \sqrt{\frac{13}{3}} \cdot t_3 \; .
                                                                  \label{eqn13}
\end{equation}

Strictly speaking, the bases of these exotic deformations are of the infinite
order. However, for the nuclear physics applications, it is important to
observe that also the rang of the spherical harmonics increases very rapidly
so that the importance of the components with the high multipolarity becomes
quickly negligible. 

%
\section{Octahedral and Tetrahedral Symmetries: Shell Structure}
\label{Sect:04}
%

In the following we are going to illustrate some characteristic features of the
single particle level spectra and of the shell structures associated with the
octahedral and tetrahedral symmetries.

%
\subsection{Octahedral Symmetry: $O^D_h$-Symmetric Single Particle Spectra}
\label{Sect:04.01}
%

\begin{figure}[t] 
\begin{center} 
\includegraphics[scale=0.39,angle=-90]{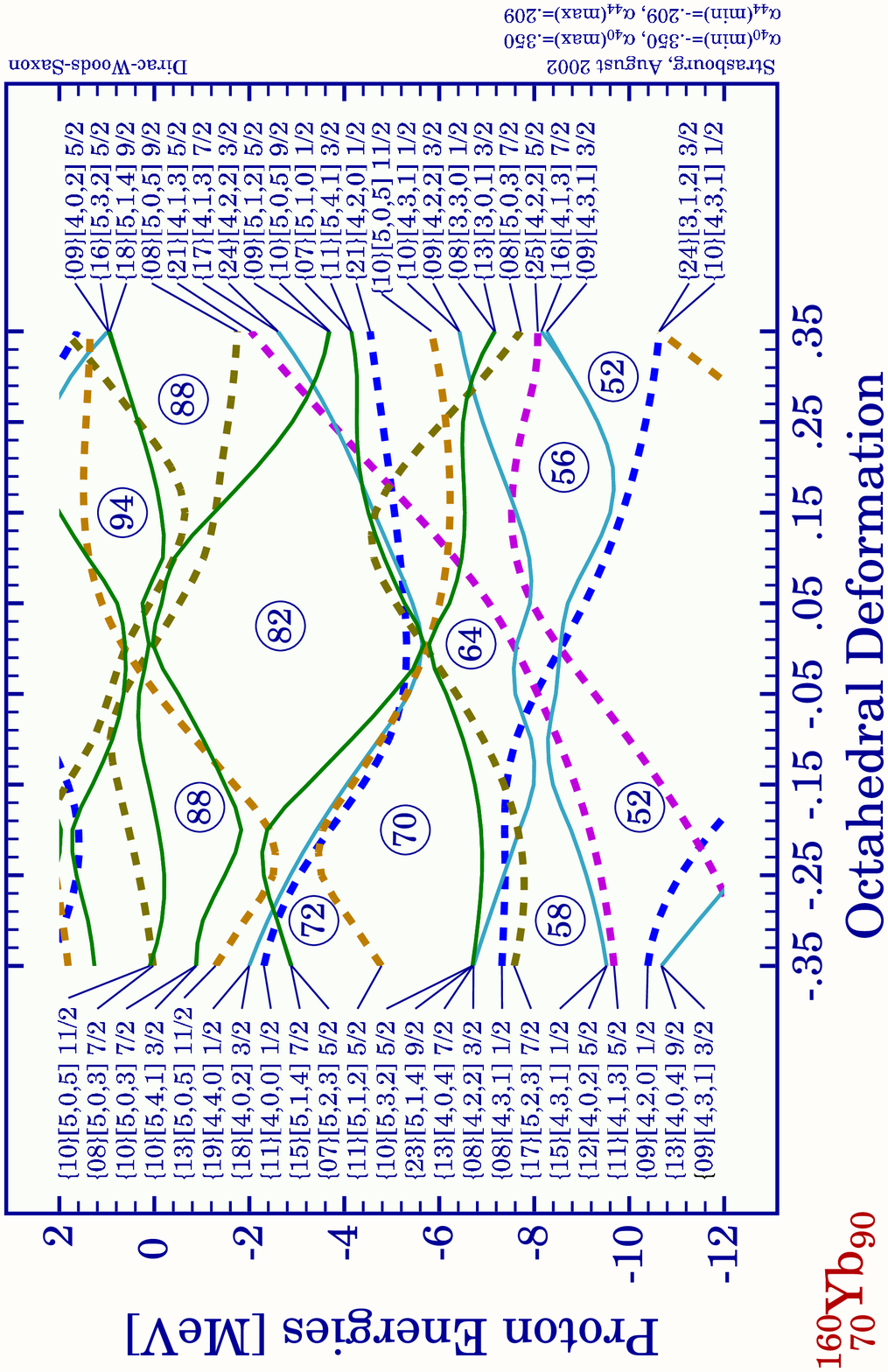} 

\vspace{-0.8truecm}
\includegraphics[scale=0.39,angle=-90]{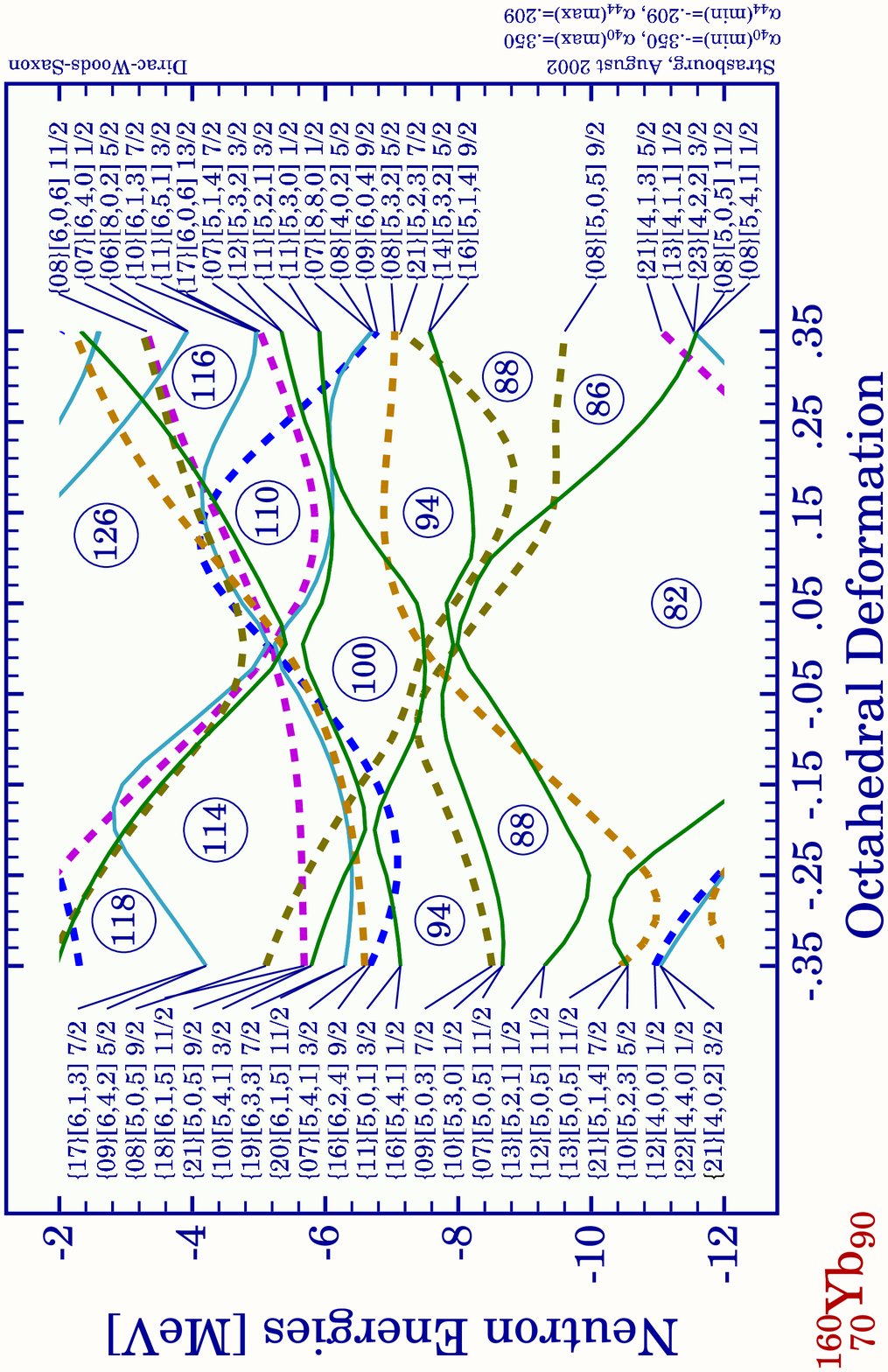}
\caption{         
         An example of single particle spectra for protons (top) and neutrons
         (bottom) valid around $^{170}$Yb nucleus in function of the octahedral
         deformation of the first order ($o_1$). The four-fold degenerate
         levels are marked with the full lines; the two-fold degenerate levels
         with the dashed lines. The curves are labeled with the Nilsson labels;
         the numbers in the curly brackets give the percentage of validity of
         each label. For further comments see the text.
         }                                  
\label{fig03}                 
\end{center} 
\end{figure}

An example of the single particle spectra corresponding to the octahedral
symmetry is shown in Fig.~\ref{fig03}. 

Let us remark first that the single particle energy curves are {\em not}
symmetric with respect to the change in sign of the octahedral deformation,
similarly as in the case of the very well known hexadecapole deformation. 

Secondly, and more importantly, let us observe that the octahedrally-symmetric
hamiltonian preserves the parity. This follows from the fact that the
corresponding shapes are modeled with the help of the spherical harmonics of
the {\em even rang} only, cf. Eqs.~(\ref{eqn08} - \ref{eqn10}). It is well
known (cf. e.g. Ref.~\cite{Cor94}) that the $O^D_h$ group has six irreducible
representations, two of them four-dimensional and the remaining four
two-dimensional. This implies that the solutions to the Schr\"odinger equation
split naturally into six families, cf. Eqs.~(\ref{eqn04} - \ref{eqn06}). From
the physics point of view it is important to notice that within each parity we
have a symmetric repartition of the irreps: one four-dimensional and two
two-dimensional ones within each parity, as it can be seen from
Fig.~\ref{fig03}. 

The very existence of the six independent classes of single-particle
wave-functions is a special feature that has not been observed so far on the
subatomic level. Any experimental evidence alluding to such a structure will be
of a great interest, and this, on the very basic level: numerous point-group
symmetries have been exploited to a large extent and for a long time in
solid-state and molecular physics and the theoretical understanding of the
underlying phenomena has profited in an important way. The mean-field theories
based on the strong interactions have not advanced towards the 'exotic'
point-group symmetries so far. 

Results in Fig.~\ref{fig03} show extremely large (over three MeV each) gaps at
Z=70 and N=114. This in itself is an important observation: the gaps of this
order of magnitude belong to the strongest known in nuclear structure physics.
Their sizes exceed e.g. the sizes of the shell-gaps responsible for the
stabilization of the superdeformed nuclear configurations. Below we will
demonstrate that these gaps may lead to static octahedral deformations in
nuclei, but our experience with the 'more traditional' gaps of this size
suggests that they  may have an important influence, among others, on the
collective oscillation properties. 

%
\subsection{Tetrahedral Symmetry: $T^D_d$-Symmetric Single Particle Spectra}
\label{Sect:04.02}
%

The tetrahedral shell structure has been studied in some detail especially
in order to establish the particularly strong tetrahedral magic gaps, cf.
Ref.~\cite{JDA02}. The corresponding tetrahedral shell closures are
predicted at:
$$ 
  Z_t = 16, 20, 32, 40, 56, 70, 90, 100, 112, 126, 
$$
for the protons, and
$$
   N_t = 16, 20, 32, 40, 56, 70, 90, 100, 112, 136,
$$
for the neutrons, showing strong similarities between the proton and neutron 
spectra.

\begin{figure}[t] 
\begin{center} 
\includegraphics[scale=0.41,angle=-90]{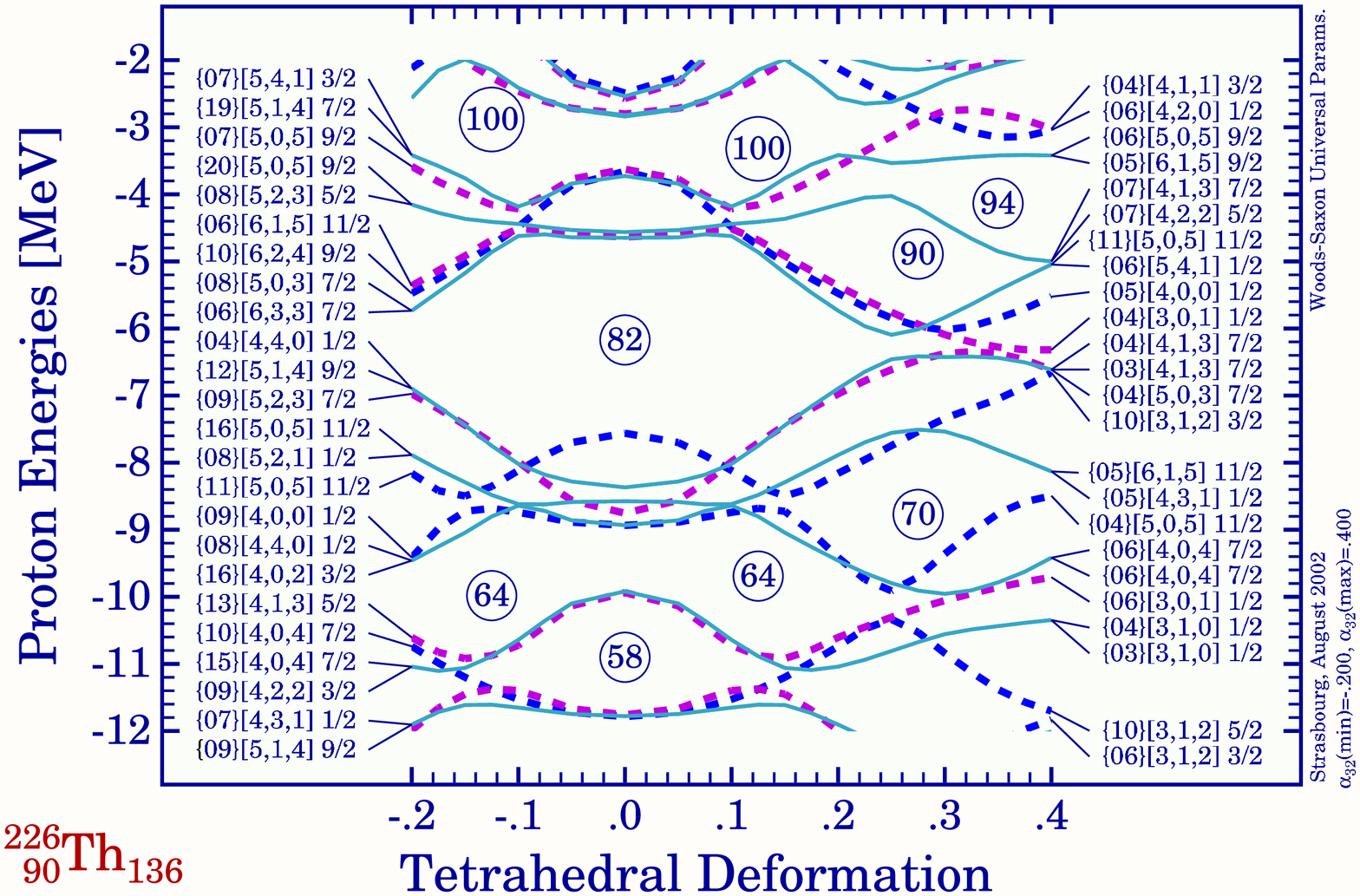} 

\vspace{-0.8truecm}
\includegraphics[scale=0.41,angle=-90]{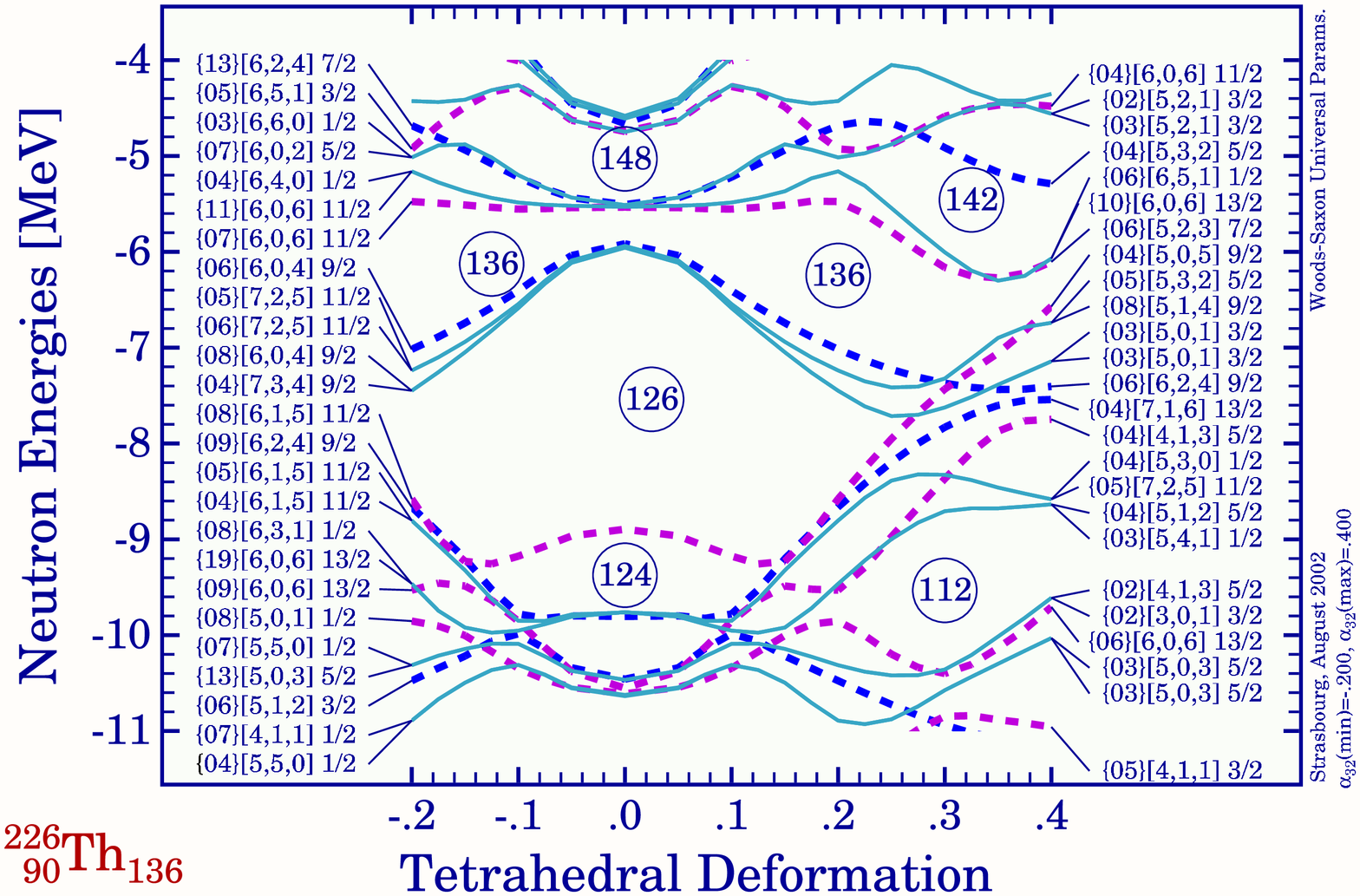}
\caption{
         Examples of the single particle proton (top) and neutron (bottom)
         spectra in function of the tetrahedral deformation of the first order,
         $t_1$. For further comments see text and also Fig.~\ref{fig03}. Large
         gaps at Z=70 and 90/94 as well as at N=112 and 136/142 deserve
         noticing.
         }                                  
\label{fig04}                 
\end{center} 
\end{figure}

Instead of over-viewing the whole series of various mass ranges we will limit
ourselves to presenting one region only. An example of the single particle
spectra corresponding to the tetrahedral symmetry is shown in Fig.~\ref{fig04}
for the nuclei in the vicinity of the $^{226}$Th. In this case the energy
curves are symmetric with respect to the change in sign of the deformation,
unlike the octahedral deformation case discussed above. Within the first order
tetrahedral deformation the corresponding tetrahedral gaps visible in the
Figure, valid for the Actinide mass range, are of the order of 2 MeV.

In principle one could think that optimal conditions to observe the
particularly stable tetrahedral nuclei would correspond to the combining of the
'doubly magic' proton and neutron configurations. Such a suggestion would have
been natural in the case of the strongest i.e. spherical shell gaps, but in the
context of the 'secondary' shell structures it is insufficient (or incorrect).
The reason is that the tetrahedral minima arise as a result of a competition
with comparably strong (or stronger) shell closures, notably at the spherical
or at the prolate deformed shapes. In such a case, a moderately strong
tetrahedral shell effect may be sufficient to produce a reasonably stable
tetrahedral minimum when {\em the competing} shell effects are weak,
non-existent, or giving rise to the positive shell energies at e.g. spherical
shapes: such a situation often takes place at certain particle numbers that are
{\em between} the traditional magic numbers corresponding to the spherical
shell closures. In the next Section we are going to illustrate this mechanism 
in some detail.

%
\section{Octahedral and Tetrahedral Symmetries: Stability}
\label{Sect:05}
%

The problem of stability of nuclei with the tetrahedral and/or octahedral
symmetries is directly related to the properties of the total potential energy
surfaces. Calculations employing the Strutinsky Woods-Saxon  method can be
considered realistic in the context and offer a priori quick and reliable
estimates. However, one needs to take into account several competing
deformation degrees of freedom\footnote{As it happens the tetrahedral
deformation of the first order (that coincides with the octupole deformation
$\alpha_{32}$) has a tendency to couple with the axially symmetric octupole
deformation $\alpha_{30}$, but other couplings cannot a priori be excluded.
Similarly, the octahedral deformation of the first order is related to the five
hexadecapole degrees of freedom and in order to obtain the realistic barrier 
estimates we need to consider up to a dozen various deformations.} in
order to avoid overestimating the heights of the potential barriers that
separate tetrahedral/octahedral minima from the other structures on the total
energy landscapes. To solve this type of a problem an algorithm has been
designed that minimizes the potential energies in an arbitrary subspace of the
multipole expansion space $\{ \alpha_{\lambda \mu} \}$. Preliminary results of
this kind of calculations have been published in Ref.~\cite{JDA02}. More
extensive calculations are in progress and here we are going to limit ourselves
to a global illustration of the shell effects on the potential energies. They
will include the macroscopic energy contributions calculated by using the most
recent version of the liquid drop model of Ref.~\cite{KPD02}.

We will illustrate the results of our preliminary calculations addressing the
problem of stability of the deformed nuclei for two discussed symmetries
separately.

%
\subsection{Octahedral Symmetry: Nuclear Potential Energies}
\label{Sect:05.01}
%

Our preliminary calculations related to the octahedral symmetry suggest
particularly strong shell effects 
\begin{figure}[h] 
\begin{center} 
\includegraphics[scale=0.35,angle=-90]{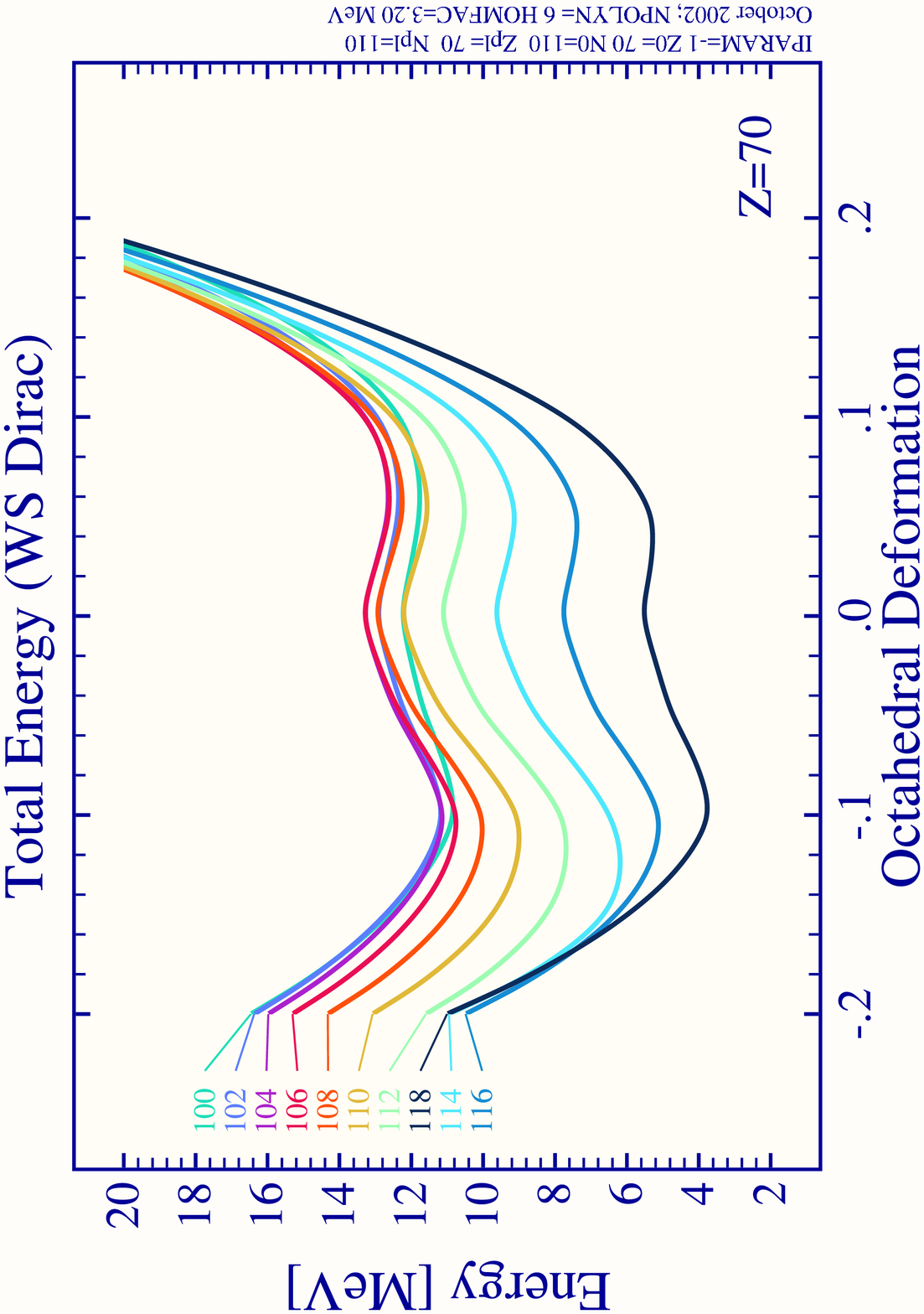} 
\includegraphics[scale=0.35,angle=-90]{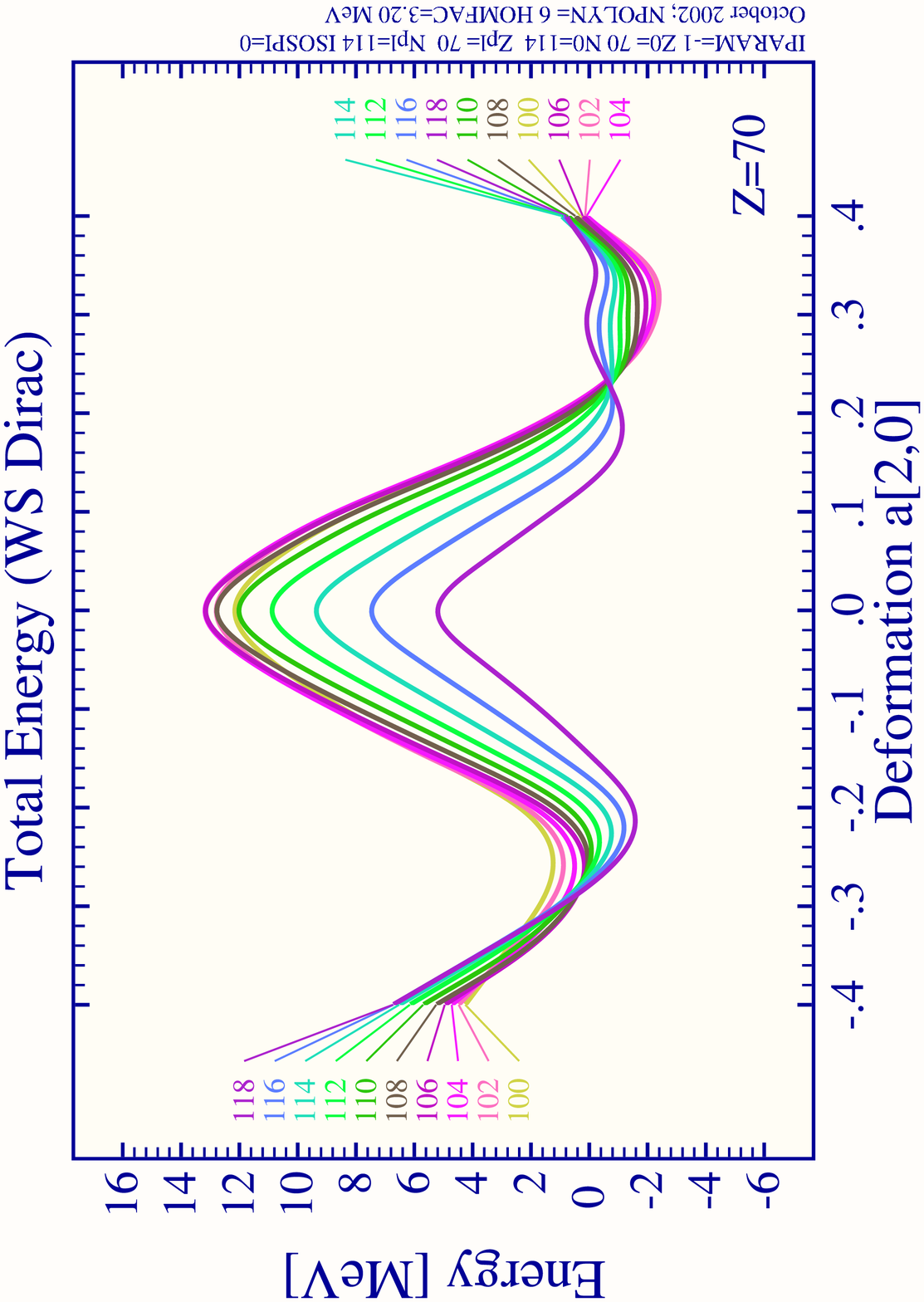}
\caption{         
         Cross-sections of the total nuclear energies in function of the
         octahedral deformation (top) and competing quadrupole deformation
         (bottom). Observe very strong effects (the energy scale is compact)
         of the octahedral deformation in {\em many} nuclei.
         }                                  
\label{fig05}                 
\end{center} 
\end{figure}
at $Z=70$ and $N=114$ with indications
to possibly weaker effect at neighboring particle numbers, cf.
Sect.~\ref{Sect:04.01}.

Since the existence or nonexistence of the octahedral minima is strongly
related to a competition between at least the quadrupole (including sphericity)
and the octahedral deformations we are going to illustrate the related effects
by comparing the simplified energy cross sections that involve these two
deformations. In Fig.~\ref{fig05} we present such energy crossections for Z=70
(Ytterbium) nuclei for several isotopes ranging from N=100 to N=118. First of
all it is worth emphasizing that the octahedral deformation susceptibility is
{\em not} a feature of the 'two magic gaps only': there are many isotopes that
manifest the minima of interest here. This follows from the fact that the shape
stability is a result of a competition among several deformation degrees of
freedom and that a minimum in a given deformation area arises either because of
the corresponding shell energies are strongly negative there or because the
shell energies are positive and particularly strong elsewhere.

The top part in Fig.~\ref{fig05} illustrates possible tendencies to produce
the octahedral-deformation minima, while the bottom part of the Figure (in
conjunction with the previous one) allows to estimate the relative excitation
energy of the exotic minimum with respect to the ground state. It can be seen
from this comparison that the excitation energies may vary, in function of the
neutron number, from about 3 MeV to about 12 MeV or so, and consequently it is
not only the size of the barrier but also the absolute excitation energy that
needs to be taken into account. More precisely, while the high barriers may be
seen as an encouraging factor that stabilizes the minima under consideration,
the high excitation energies can be seen as a direct measure of difficulties
with the population of those states, the larger the energy the more difficult
the population. 

The nuclei with neighboring Z-values provide a priori several further
candidates for the octahedral symmetry studies. At this stage we may conclude
that the octahedral 'susceptibility' is not limited to a couple of nuclei only
as the single-particle diagrams presented earlier may suggest. But to draw the
definite conclusions it will be necessary to complete the more involved,
multidimensional potential energy calculations.

%
\subsection{Tetrahedral Symmetry: Nuclear Potential Energies}
\label{Sect:05.02}
%

\begin{figure}[t] 
\begin{center} 
\includegraphics[scale=0.37,angle=-90]{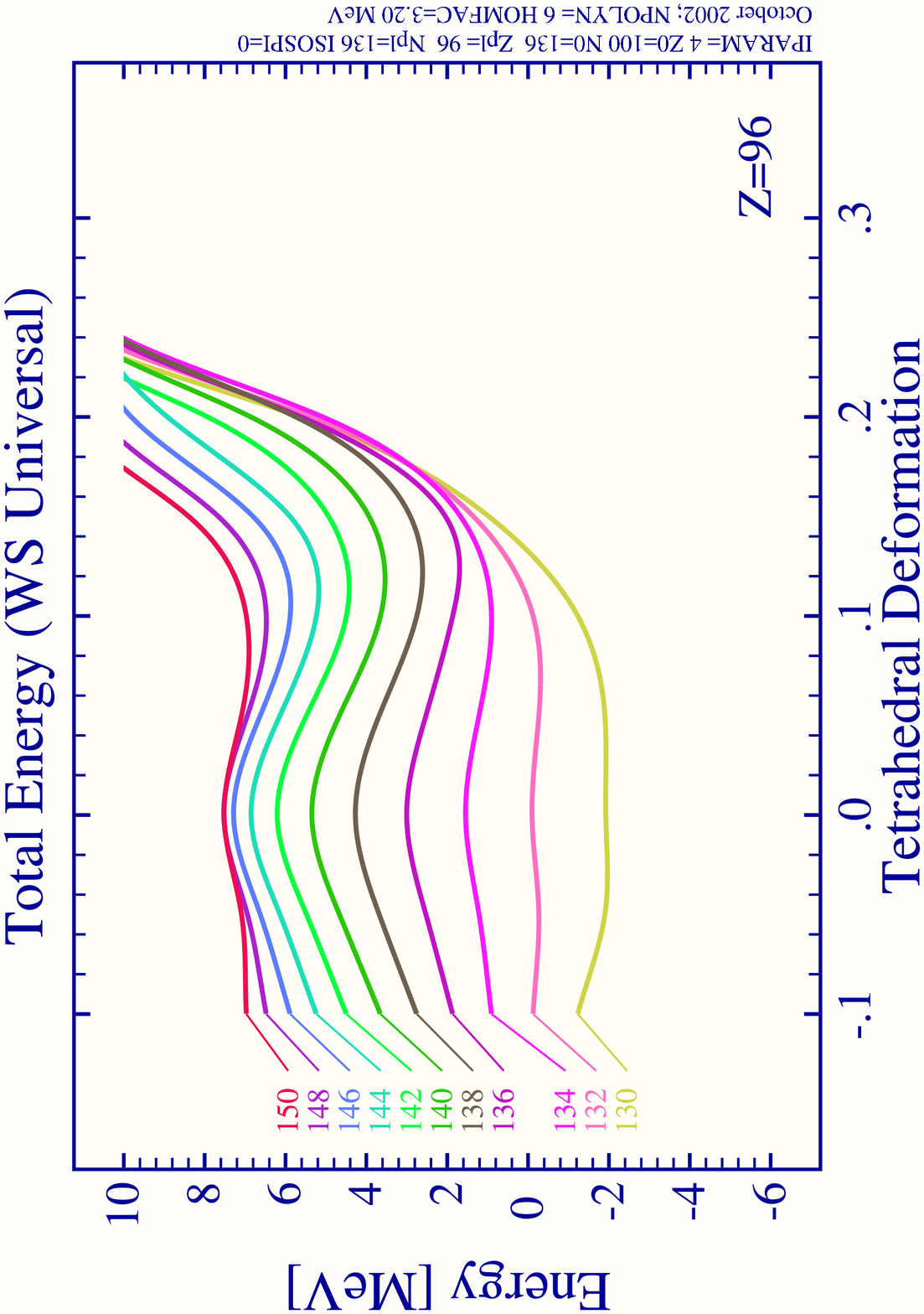} 
\includegraphics[scale=0.37,angle=-90]{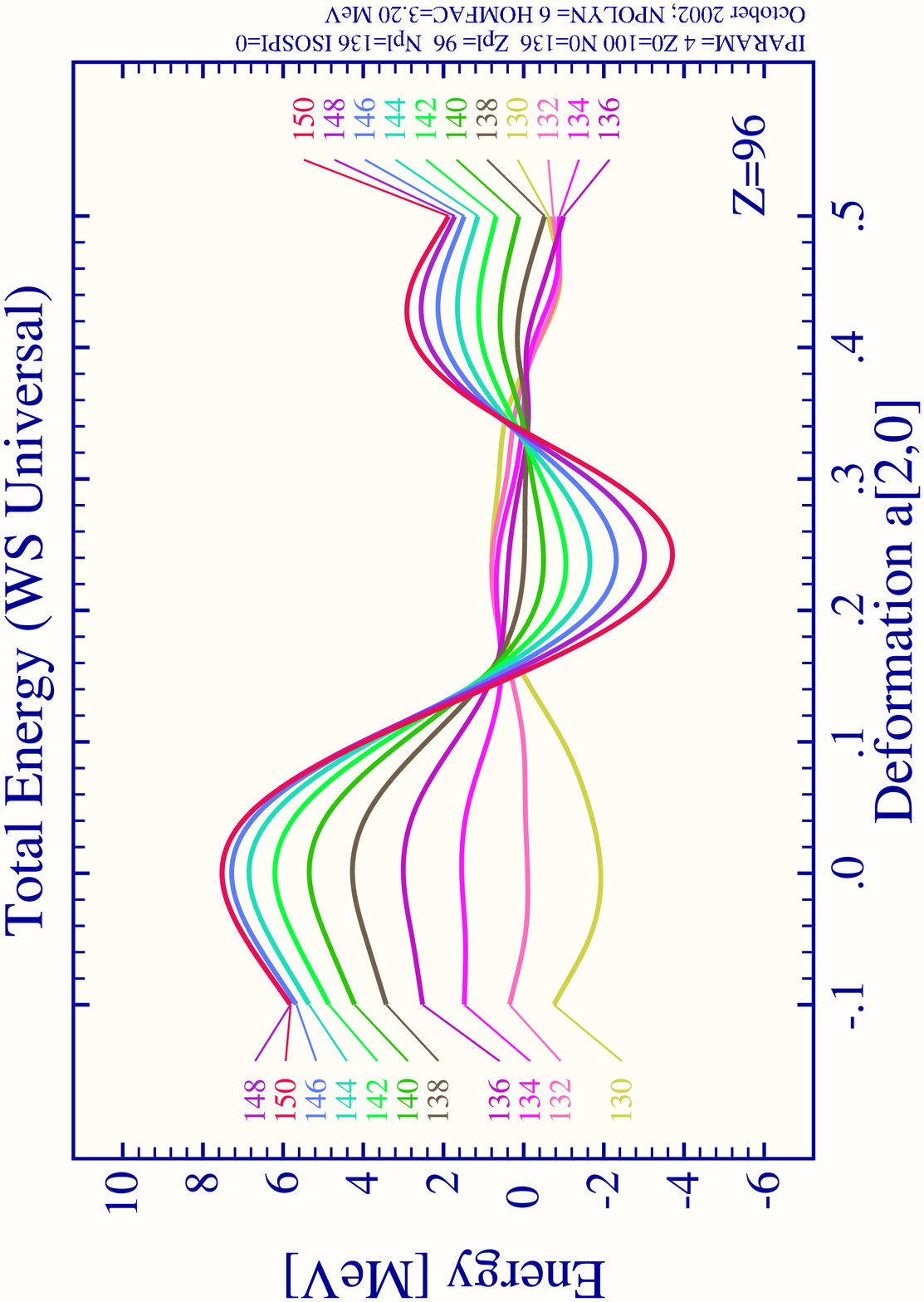}
\caption{         
         Cross-sections of the total nuclear energies in function of the
         octahedral deformation (top) and competing quadrupole deformation
         (bottom). For further comments see text and the preceding Section.
         }                                  
\label{fig06}                 
\end{center} 
\end{figure}

A global 'first-test' analysis aiming at a comparison of the susceptibility
towards the ocurrence of the tetrahedral shapes among many nuclei has been
performed in analogy to the one presented above for the case of the octahedral
symmetry. Here we select only one nuclear range focusing at the masses roughly
between 230 and 250. The corresponding total potential energy cross-sections
are shown in Fig.~\ref{fig06} for illustration; the form of this representation
follows the one in Fig.~\ref{fig05}.

It can be seen from the Figure that there are groups of nuclei susceptible to
form the tetrahedral-deformed minima, yet the excitation energies related to
those highly symmetric structures may be very high. Indeed, a comparison
between the results related to the energy minima in the top part of
Fig.~\ref{fig06} with the absolute energy minima of the curves in the bottom
part of the Figure indicates that the energy differences may vary between
(2.5 - 3) MeV and about 10 MeV or so. We conclude that an optimal choice
for experimental test will require a compromise between the barrier
heights (possibly large) and the excitation energies (possibly low).  

%
\section{Nuclei with Tetrahedral Symmetry: Population, Decay and Observation 
         Possibilities}
\label{Sect:06}
%
We believe that the hypothetical tetrahedral structures should be more abundant
as compared to the octahedral ones and thus we would like to focus the
following discussion on the former rather than on the latter.

In addressing the problem of an experimental discovery of the discussed nuclear
symmetries one may first ask a question as to why none of them has so far been
observed? Obviously one may advance several hypotheses ranging from the most
pessimistic ones ('since they have not been seen so far, perhaps they do not
exist') to the most optimistic ones ('the evidence has been already collected,
the affirmative answers are stocked on the tapes with the results of already
performed experiments - the only problem: nobody has thought about it').

Let us try to discuss this very basic question first: why do we believe in the
existence of the new symmetries on the sub-atomic level? The arguments in
favor of the existence of these symmetries are based today exclusively on the
theory grounds. The present day nuclear microscopic theories are advanced and
performant tools 'tuned' to describe the experimental results in numerous
sub-fields of our domain. In particular the self-consistent Hartree-Fock
methods have reached a high level of predictive power, especially since 
artificial constraining conditions imposed in the early stages of the
development to simplify the computing - are nowadays removed (cf. e.g. Ref.
\cite{DDu01} and references therein). In the present context it has been
verified in a few cases (a more systematic study is in progress) that the
Hartree-Fock iterative procedures started at initial configurations with all
symmetries totally broken may converge to the highly symmetric solutions of the
type discussed in this paper and moreover, in accordance with the Z and N
numbers for which strong shell effects have been predicted. In particular, the numerical coefficients that
define the proportions of various multipoles [Eqs.~(\ref{eqn08}-\ref{eqn10})
for the octahedral symmetry and Eqs.~(\ref{eqn11}-\ref{eqn13}) for the
tetrahedral symmetry] predicted by the group theory are reproduced by the
self-consistent Hartree-Fock solutions down to the computer accuracy. One can
hardly consider this type of the result accidental. We believe that the 
remaining question is not: {\em Whether} but rather: {\em On what level of 
probability} the configurations under discussion can be populated and detected?  
\begin{figure}[t]  
\begin{center} 
\includegraphics[scale=0.45]{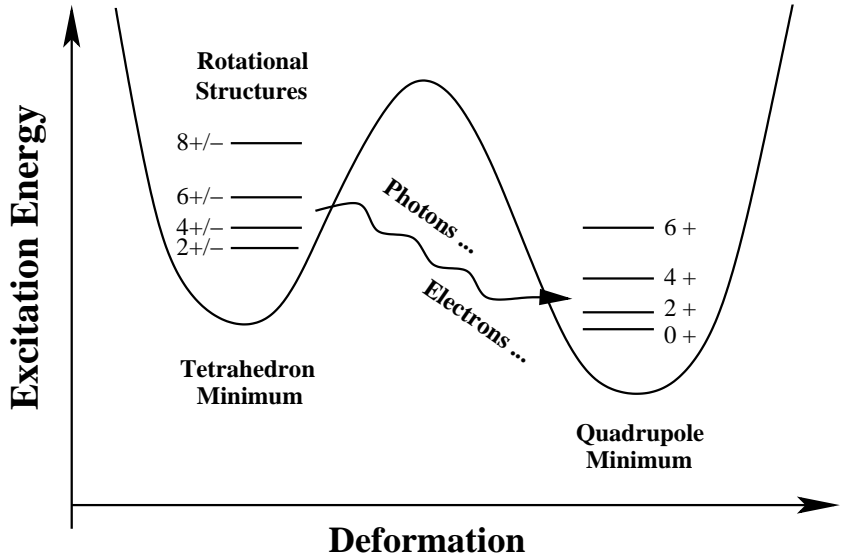} 
\caption{                                   
        Schematic illustration: structure and possibilities of the decay out of
        a tetrahedral minimum. Since the lowest-order tetrahedral deformation
        has the same geometrical features as the octupole deformation
        $\alpha_{32}$, the concerned nuclei may generate parity-doublet
        rotational bands known from the studies of the octupole shapes.
        Establishing the structure of the bands (parity doublets?), the nature
        of the inter- and intra-band transitions (dipole? quadrupole?
        octupole?), the properties of the side-feeding and the decay branching
        ratios - all that will greatly help identifying the symmetry through
        experiments. 
        }
\label{fig07}                  
\end{center}  
\end{figure}

Preliminary calculations based on the Dirac-Strutinsky calculations give not
only the prediction of the very existence of the tetrahedral (octahedral) {\em
shell-structures} but also several indications about the related {\em
spectroscopic properties}. According to those calculations the minima on the
potential energy surfaces that correspond to tetrahedral (octahedral) shapes
are surrounded by potential barriers whose heights vary between a few hundreds
of keV up to a couple of MeV. The excitation energy associated with these
minima may vary typically between $\sim 1$ MeV and several MeV (cf. preceding
Section). It is therefore possible that the minima of interest will lead to
isomeric structures analogous to the ones associated with the prolate-oblate
shape coexistence known e.g. in the Mercury region.

\begin{figure}[t] 
\begin{center} 
\includegraphics[scale=0.35]{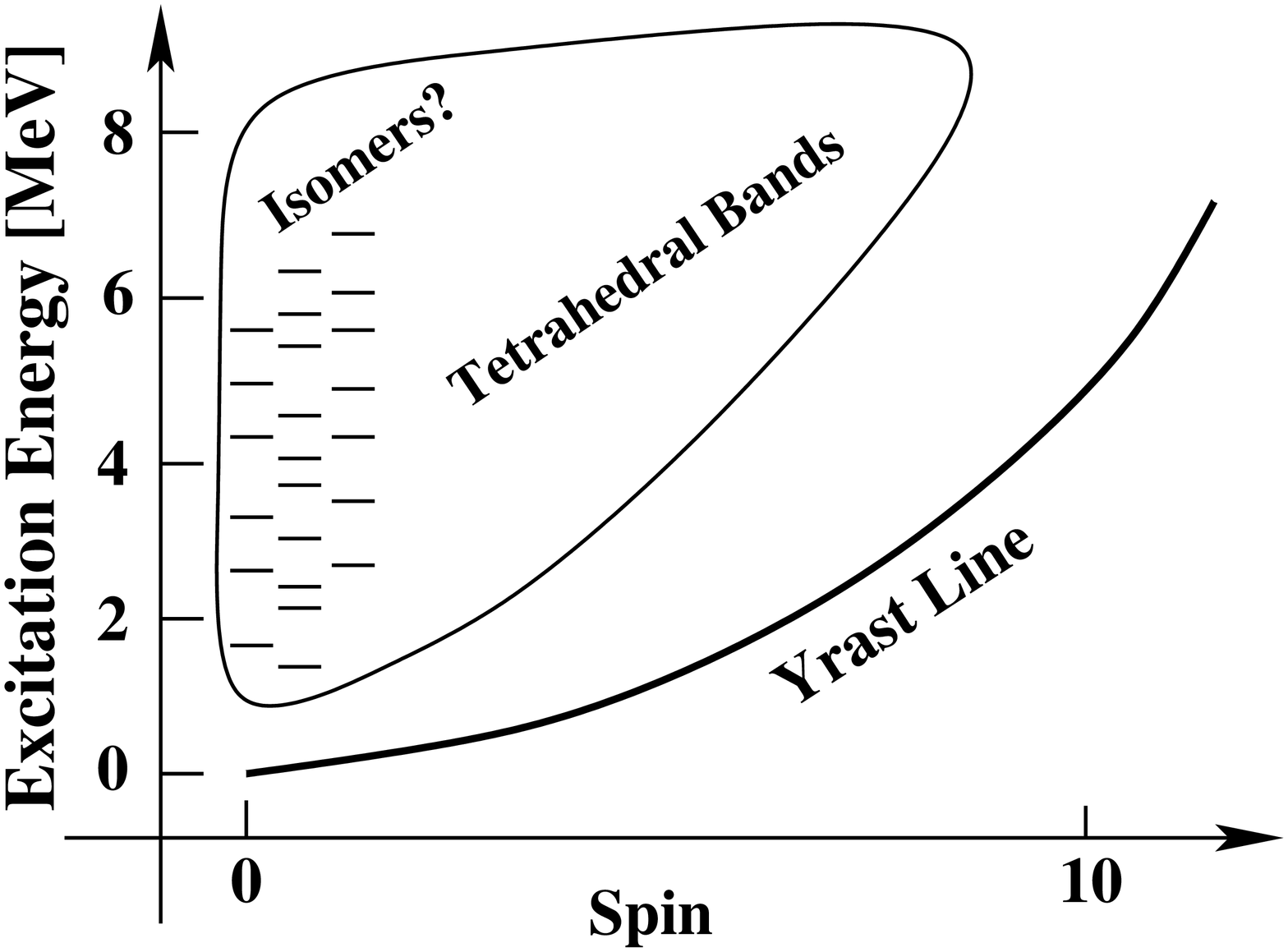} 
\caption{
         A schematic 'phase diagram' illustrating expected positions of the
         hypothetical low-spin isomers in various nuclei as well as the bands associated
         with tetrahedral minima (for some quantitative examples cf. preceding
         Sections). One may expect that in some nuclei the band-heads (isomers?)
         may lie prohibitively high in energy, while in some others [ (1-3) MeV
         above the ground state] they may be much easier to populate. The
         scales of the energy and of the spin are realistic.
         }
\label{fig09}                 
\end{center} 
\end{figure} 

Unlike yrast-trap isomers, whose configurations are related to the
particle-hole excitations with the spins nearly parallel to the symmetry axis,
the configurations associated with tetrahedral (octahedral) excited nuclei
correspond to deformations with no symmetry axes. Therefore the nuclei in
question may always decay through rotational transitions down to the band-head,
cf. Fig.~\ref{fig07}. The yrast trap configurations leading to isomers may in
principle appear at any spin ranging from 0 to $\sim$ 30 $\hbar$, possibly
higher. The isomers associated with new symmetries could be expected mostly at
the lowest spins of the order of $I$ $\sim$ (0 to 2)~$\hbar$, at or near the
band-heads.

The new symmetries are expected to be a relatively low spin, but possibly high
excitation-energy phenomenon. In reference to Fig.~\ref{fig09} we may conclude
that the main difficulty is likely to be associated with the population rate of
states that lie low in spin, say $I \sim (0 - 14)$ $\hbar$, but whose
excitation energies correspond to $(1 - 4)$ MeV or more. Consequently,
reactions with light projectiles may be the first choice there. According to
such a scenario the target nuclei selected will lie close to the
$\beta$-stability line so that the final nuclei 'to start with' will most
likely also belong to that area.

Obviously a possible use of the radioactive beams will be another source of
challenges.  

\begin{figure}[t] 
\begin{center} 
\includegraphics[scale=0.45]{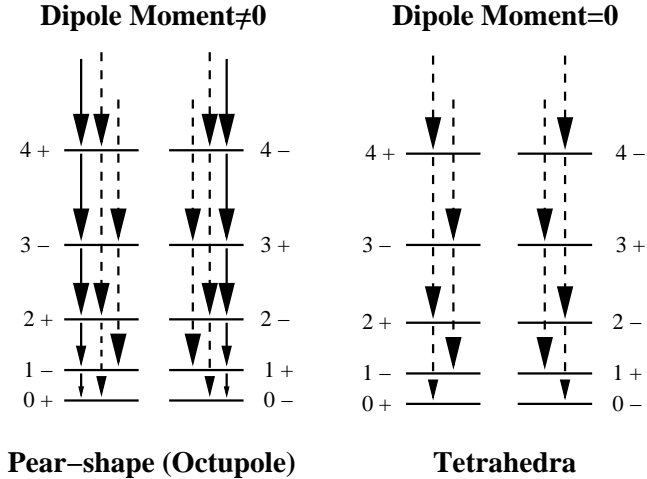} 
\caption{        
         Schematic illustration: rotational bands associated with the
         tetrahedral shapes (right) as well as those associated with the
         axially-symmetric octupole (pear) shapes (left) are expected to
         preserve the simplex quantum number and, in the extreme limit, to
         produce the degenerate parity-doublet bands. The tetrahedral shapes,
         unlike the pear-shapes, generate vanishing dipole moments. Quadrupole
         transitions are marked with dashed-, dipole transitions with
         full-lines; octupole transitions are possible in both cases but are
         not marked (see the text).
         }
\label{fig10}                 
\end{center} 
\end{figure} 

To establish an as simple as possible a scenario of reference, let us begin
with a model of an {\em ideal} situation, according to which our hypothetical
nuclear configuration is associated with a strong tetrahedral minimum. Suppose
that we can neglect any coupling to vibrations (such as e.g. the zero-point
motion) and/or shape-polarizations that could possibly contaminate the purity
of the discussed symmetry. At this limit, both the dipole and the quadrupole
electromagnetic transitions are strictly speaking zero and the first allowed
ones correspond to $\lambda = 3$ (octupole). Within such an ideal scenario the
decay spectra of the tetrahedral nuclei will be composed of {\em octupole
transitions only} while preserving other rules characteristic of the
conservation of the simplex quantum number\footnote{Recall that the simplex
transformation, say ${\mathcal{S}}_y$, is by definition equal to the product of
the $180^o$ rotation about the $\mathcal{O}_y$-axis, ${\mathcal{R}}_y$, and the
inversion $\mathcal{I}$ and we have ${\mathcal{S}}_y = {\mathcal{R}}_y \cdot
\mathcal{I}$. In an ideal case of the strong, well defined minimum the energies
of the inter-band dipole transitions correspond to half of the related
quadrupole intra-band transitions.} - at variance with the illustration in
Fig.~\ref{fig10}. In other words: an extreme beauty of the underlying physical
picture consists in the fact that the rotational energy levels satisfying, at
least to a leading order, the usual $I(I+1)$-rule would be connected through
the $(\Delta I = 3)$-octupole transitions rather than through the usual
quadrupole ones.

In realistic situations, various deviations from the ideal limit are to be
expected. In fact, in Fig.~\ref{fig10} we have chosen to illustrate what we
believe is a more likely situation, namely, the presence of a quadrupole
polarization and/or a 'residual' quadrupole deformations 'contaminating' the
pure symmetry. In such a case the quadrupole transitions accompanying the decay
of a tetrahedral nucleus are expected to be weak but most likely dominating the
octupole transitions. In contrast, the pear shape nuclei known today have their
octupole deformations superposed with the quadrupole ones, and their spectra
are characterized by strong quadrupole and dipole transitions at the same time.
(One may conjecture that, given the mass of the compared nuclei, the quadrupole
transitions in tetrahedral nuclei could be at least one order of magnitude
weaker than those in the normally deformed octupole ones).

Another mechanism that is likely to modify the ideal model situation is a
possibility that the parity doublet structures are split by a considerable 
fractions of their energy. Such a mechanism is likely to occur when the
tetrahedral minima are not sufficiently deep. As a consequence, instead of
having a particularly simple picture as the one in Fig.~\ref{fig10}, the bands
appearing degenerate there will be considerably displaced with respect to one
another.

The decay of the lowest energy states corresponding to the tetrahedral minima
may be of particular interest. The corresponding energies of excitation with
respect to the ground-state varying between numbers slightly over 1 MeV up to a
couple of MeV, the internal pair production may provide an extra signal. Given
the fact that mean radius expectation values, $< r^2 >$, are likely to differ
considerably between the tetrahedral excited- and the quadrupole ground-state
minima, the $E0$ transitions connecting those states are likely to be
enhanced.


\section{Towards First Principles: Nuclear Quantum Mechanics}
\label{Sect:07}


It turns out that the symmetry-induced properties of the nucleonic levels and
wave functions, as suggested in this paper, are unprecedented in nuclear
structure studies and provide interesting new challenges already at the level
of the nuclear quantum mechanics. First of all, an existence of new quantum
numbers is predicted. One of them may take three values in the case of the
tetrahedral symmetry; the other two, each of which take in turn three values
possible, in the case of the octahedral symmetry, cf. Table \ref{tab01}. These
good quantum numbers will replace the well known {\em  signature quantum
number} in the case of the parity-preserving octahedral shapes, and the {\em
simplex quantum number} in the case of intrinsic-parity breaking tetrahedral
deformation. 

\renewcommand{\arraystretch}{1.5}

\begin{table}[ht]
\begin{center}
\small
\caption[T1b]{\label{tab01} 
              CHALLENGES ON THE LEVEL OF QUANTUM MECHANICS\\}
              \vspace{+0.2truecm}
              \begin{center}
              (Unprecedented Quantum Features Related to $T^D_d$ and
              $O^D_h$ Nuclear Symmetries)
              \end{center}
              \vspace{+0.2truecm}
                          
\begin{tabular}{||c||c|c||c||}
\hline
\hline
Properties & \multicolumn{2}{c}{ High Symmetries }     \vline \vline & 
                                 'Usual' Symmetries    \\
\hline
(or observables) & Tetrahedral & Octahedral & $D^D_2$ ('tri-axial')  \\
\hline
\hline
No. Sym. Elemts. & 48      &  96 &  8  \\
 Parity         & NO      & YES & YES  \\
 Degeneracies        & 4, 2, 2 & $\underbrace{4,2,2}_{\ds \pi=+} \;\;$
 $\underbrace{4,2,2}_{\ds \pi=-} $&  $\underbrace{2}_{\ds \pi = +} \;\; 
 \underbrace{2}_{\ds \pi = -}$\\
 Quantum Numbers      & 3       & $3 + 3 + 2\,(\pi= \pm) $&  2 $\,(\pi = \pm )$  \\
\hline
\hline
\end{tabular}					     
\end{center}					     
\end{table}

In Table \ref{tab01} we compare some properties, parameters or observables
related to the specificity of the $T^D_d$ (tetrahedral) and $O^D_h$
(octahedral) double point-group symmetries. In particular, it is easy to see
that the numbers of symmetry elements in the case of $T^D_d$ or $O^D_h$ exceed
by important factors the number of symmetry elements associated with the well
known, 'standard', triaxial-symmetry type shapes. There are in total six
families of the single-particle levels that can be associated with solutions to
the $O^D_h$-symmetric hamiltonian; three of them  belong to the positive- and
three to the negative-parity and, moreover, one such a family within each set
is characterized by the four-fold degeneracy. A similar property holds for the
$T^D_d$-symmetric hamiltonians except for the parity that is not conserved in
this case. The last line in the Table shows how many values can the discrete
quantum numbers take in relation to the symmetries compared.

Another important aspect will need to be considered: the tetrahedrally- or
octahedrally-symmetric nuclei, if present in nature, are expected to exhibit
the collective rotational properties that are very different from what we have
learned by studying the rigid tri-axial rotors. First of all, let us remind the
reader that the classical moments of inertia associated with a
tetrahedrally-symmetric object are all three equal. As a consequence, the usual
way of treating the collective rotation will give the same result as in the
case of a rotating rigid {\em sphere!} (if we accept the rigid-rotation model)
or infinity (in the case of the superfluid-rotation models where we impose
rotation about a symmetry axis). Yet: a tetrahedrally-deformed nucleus has
clearly no symmetry axis, its orientation in space can be very well defined and
it is to be expected that two nuclei that differ in the {\em size} of the
tetrahedral deformation will also have non-trivially different rotation-energy
spectra.

It then becomes clear that to study such objects we will need to give-up the
traditional rotor hamiltonian expressions that are quadratic in spin: the
tetrahedral symmetry can be modeled with the help of polynomial expressions of
{\em at least third order}. But then we are confronted with another beautiful
problem both from the mathematics and physics points of view that can be
introduced as follows. Consider an even-even nucleus whose rotational spectra
are expressed in terms of integer spins (i.e. 'boson-type', as opposed to
'fermion-type' half-integer spins, in the case of odd-A nuclei). The
corresponding rotor hamiltonian has therefore the 'usual' tetrahedral group
$T_d$ as its symmetry group. At the same time the Schr\"odinger equation that
governs the motion of the nucleons (fermions) in the same nucleus is invariant
under the \underline{\em double} point group symmetry $T^D_d$. From the
mathematics point of view these two groups are totally different, they have
different numbers of groups elements as well as of the equivalence classes and
thus of the numbers and types of irreducible representations. In particular:
both classical (as opposed to double) point-groups, $T_d$ and $O_h$, have the
same number of {\em five} irreducible representations, two of them
one-dimensional, one two-dimensional, and two three-dimensional. These
properties imply an unprecedented degeneracy patterns in terms of the
collective rotational spectra: these degeneracy patterns, already exotic,
do not resemble at all the exotic degeneracy patterns predicted earlier in the
article for the single-nucleonic spectra. (Some mathematical aspects related
to quantum mechanical features of highly-symmetric rotating objects are
discussed in Ref.~\cite{AGD03}).

This brings us to another aspect of the quantum description of the problem: for
the first time we will be forced to consider the three-dimensional aspect of
the quantum rotation from the start! - no simplifications such as alignments on
the 'principal axis' (no such axis can be defined) will be possible. 

All these unprecedented quantum mechanisms will (perhaps) not be very easy to
establish in experiment. But if confirmed experimentally, the presence of these
new symmetries on the subatomic level will strongly influence our present-day
understanding of nuclear phenomena.  


\section{Summary and Conclusions}
\label{Sect:08}


In addition to the general, new quantum mechanics related features summarized
in the preceding Section, the presented approach involves a new way of thinking
about the nuclear stability: it is, for the first time\footnote{For a long
time, the spherical-symmetry considerations have defined a standard when
studying the symmetries in all domains of physics, in particular in relating
the problems of degeneracy of levels to the problem of increased stability of
the related configurations. Here we return to this very basic quantum
mechanical problem in a non-trivially new physical situation generated by the
possible ocurrence of the tetrahedral and octahedral symmetries in nuclei.}
based on the genuine symmetry considerations, and {\em not} e.g. on comparing
the harmonic oscillator axis-ratios: 
$$ 
  a : b : c 
  = 
  {\rm ratios~of~small~integers} \; . 
$$

Another aspect of novelty consists in going away from the multipole expansion,
that has been a standard approach during a long time, for instance when
calculating the nuclear potential energy surfaces [ e.g. $(\beta,\gamma)$-plane
representation in terms of $Y_{20}$ and $Y_{22}$ multipoles, hexadecapole
$Y_{40}$-deformation etc.]. In this paper we have shown how to construct the
point-symmetry oriented bases instead of the spherical harmonic basis, the
latter remaining of course an ideal tool for studying the $SO(3)$-symmetry
related properties. 

We would like to turn the reader's attention to the fact that if we include the
possibility of existence of the tetrahedral and/or octahedral symmetries, an
important sub-field of our domain that deals with the Shape Coexistence
Phenomena, will contain from now on an impressive number of configurations to
study, many of them possibly in one single nucleus (!) These are: Prolate,
Spherical, Oblate, Triaxial in various forms, Tetrahedral, Octahedral,
'Triangular' ($\mathcal{C}_3$), Octupole, Superdeformed, Hyperdeformed, and
possibly more nuclear shapes and related symmetries.

So far in the nuclear structure physics we were contenting ourselves with the
presence of two types of nucleonic level degeneracies: either the (2j+1)-fold
degeneracy associated with the nucleonic levels in a spherically-symmetric
potential or the double (spin-up, spin-down) time-reversal degeneracy of
Kramers in the case of deformed nuclei. Since we were dealing with these
degeneracies for a long time being confronted with them in nearly all
microscopic models developed so far, their presence has evolved to a kind of a
'trivial property'. In the problem presented here we deal for the first time
with what we would thus call 'non-trivial' degeneracies of the nucleonic
levels.

It remains to be seen, as one of important items on the challenge list, whether
one will be able to talk about analogies when comparing to the molecular
symmetries based on the forces of the infinite range. These are {\em opposed}
to the short range strong-interaction nuclear-forces that, under some specific
circumstances related to the nuclear shell structure, may possibly generate the
same symmetries. This problem touches upon the very basic issues in nuclear
structure physics directly related to the concepts of the mean-field and of 
spontaneous symmetry breaking.

%

%

%
\end{document}